# Comparing the Ag-content of poltinniks using X-ray fluorescence

S Ferguson and S Williams

Department of Physics and Geosciences, Angelo State University, San Angelo, Texas USA

E-mail: scott.williams@angelo.edu

**Abstract**

X-ray fluorescence experiments have been performed in order to analyze the elemental composition of four Russian 50-kopek coins ("poltinniks") minted during 1913, 1921, and 1924. By comparing the intensities of the Ag $K_\alpha$ X-rays emitted by the poltinniks, we were able to determine whether the Ag-content of the coins were equal. One of the goals of this study was to determine whether or not legislation was carried out that required the proportions of Ag and Cu used in the minting of coins in 1924 to be identical to those minted in previous years. Also, the intensities of the Ag $K_\alpha$ X-rays emitted by 1924 poltinniks minted in London and Leningrad were compared. Our results suggest that the percent difference in the proportions of Ag present in each of the coins is no more than 5.5%.

## 1. Introduction

*1.1. Educational Objectives*

When designing research projects for undergraduate students, it is important to try to ensure that the research strengthens the students' basic understanding of the physical phenomena involved in the research while also providing students with practical experience working with equipment typically found in industrial and university laboratories. Experiments such as the ones described in this report allow undergraduate students to familiarize themselves with several different radiative processes including bremsstrahlung emission, Compton scattering, and characteristic X-ray emission (due to both photoionization and electron impact). These types of experiments also give students experience working with equipment such as radiation detectors, various electronics such as multichannel analyzers, and spectroscopy software. Furthermore, experiments such as the ones described in this paper can help to introduce undergraduate students to the safety precautions that are necessary when working with ionizing radiation.

     The elemental composition of materials can be determined using X-ray fluorescence (XRF). XRF analysis of a sample involves incident radiation (typically X-rays or gamma rays) exciting or ionizing the atoms of the sample. When the atomic electrons "relax" and transition to lower-energy states, characteristic radiation is emitted. Analysis of the radiation spectrum can help determine the elemental composition of the sample. The experiments described in this report involved comparing the intensities of the Ag $K_\alpha$ X-rays emitted by Russian 50-kopek coins (often referred to as "poltinniks") minted in 1913, 1921, and 1924 when they were irradiated with bremsstrahlung produced by 40-keV electrons incident on a thick Au target. The intensities were studied in order to compare the Ag-content of each of the poltinniks.

*1.2. Historical Background*

The elemental compositions of four poltinniks were analyzed using XRF in our experiments. The first poltinnik analyzed was minted in 1913, a few years before the October Revolution of 1917, during the reign of Nicholas II. The second poltinnik analyzed was struck in 1921, after the formation of the Russian Soviet Federative Socialist Republic. The third and fourth poltinniks that were analyzed were both minted in 1924; one of the poltinniks was struck in Leningrad, while the other one was struck at the Royal Mint in London. Figure 1 shows photographs of the four poltinniks used in the experiments.

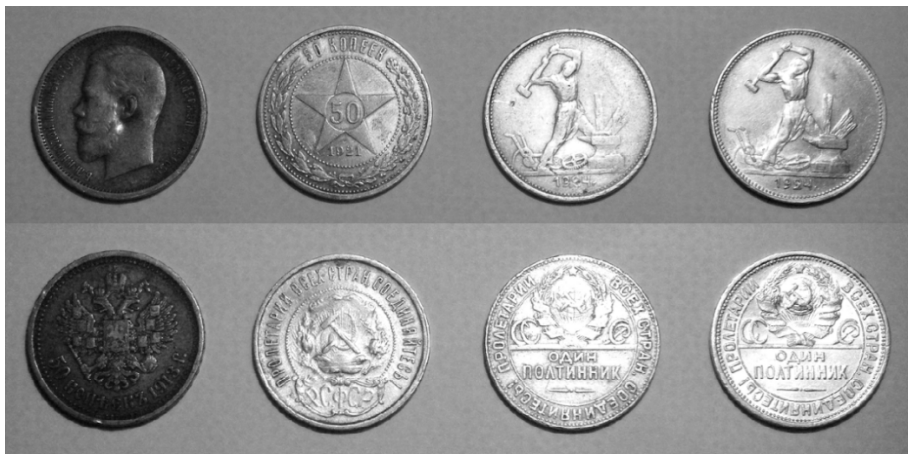

**Figure 1.** Photographs of the poltinniks minted in (from left to right) 1913, 1921, 1924 (Leningrad), and 1924 (London).

Early in 1924, a series of legislation related to currency reform was passed. "The third legislative enactment in the series of currency reform was the joint Decree of the Central Executive Committee and the Council of the People's Commissaries dated February 22nd, dealing with the mintage and issuing of silver and copper coins. […] The Government thought it expedient to retain the former proportion of pure [Ag] and [Cu] in the coins and to make no alterations in respect of their weight, diameter or any other details" [1]. One of the goals of the experiments described in this report was to determine whether or not the legislation was carried out by comparing the XRF spectra produced by the 1924 poltinniks to the spectra produced by the 1913 and 1921 poltinniks.

During the early 1920s, the Soviet Union faced a "small-change crisis" [2] and the Central Executive Committee requested that the mint in Leningrad increase production of coins. However, they were unable to produce the amount needed in sufficient time and the Royal Mint in London began to mint coins for the Soviet Union, as well. The Soviets ordered 40 million copper piataks (5-kopek coins) and 40 million silver poltinniks from the Royal Mint [2]. The two poltinniks struck in 1924 were analyzed using XRF to determine whether or not the proportions of Ag in the coins are the same.

**2. Experimental Procedure**

XRF analysis was performed using an Amptek Mini-X X-ray generator and an LD Didactic GmbH Si-PIN detector. The Mini-X produces bremsstrahlung by accelerating electrons toward a thick Au anode. All XRF analysis was performed using an X-ray beam collimator, an accelerating voltage of 40 kV, and an electron beam current of 75 µA. Data was collected over a "live" time of 1200 seconds for each analysis. The orientations of the Mini-X, poltinniks, and Si-PIN detector were the same in all experiments. The Si-PIN detector was positioned at a forward angle of approximately 160° from the X-ray beam line. Both sides of each poltinnik were analyzed and the XRF spectra produced by both sides of each coin were summed (and are henceforth referred to as a single spectrum).

Before performing the experiments, it was necessary to calibrate the energy scale. This was done using several 99.99% pure samples purchased from Alfa Aesar, USA. Energy calibration is achieved by examining the XRF spectrum produced using a sample of known elemental composition and determining what channels correspond to what energies. For example, a pure Ag sample produced a spectrum with a prominent peak centered at channel 764. This peak is associated with ~22.2 keV Ag $K_\alpha$ X-rays. Using this information and XRF spectra produced by other samples, the energy scale was calibrated.

While it is a common practice to treat the surface of objects before XRF analysis [3], we did not treat the poltinniks used in our experiments due both to curatorial concerns and to the fact that certain cleaning treatments have been shown to reduce the overall weight of coins [4].

## 3. Results and Discussion

Figure 2 is the XRF spectrum produced by the poltinnik minted in 1913. The peaks associated with the Ag $K_\alpha$ and $K_\beta$ X-rays (with energies of approximately 22.2 and 24.9 keV, respectively) can clearly be seen in the spectrum along with several other peaks. Numismatic publications [5, 6] suggest that all of the poltinniks analyzed in our experiments are 90% Ag by weight. The peaks at approximately 8.1 and 8.9 keV are associated with Cu $K_\alpha$ and $K_\beta$ X-rays, respectively, which suggests that the remaining 10% of the poltinnik's mass is composed primarily of Cu (the Cu $K_\beta$ peak is not labeled in the figure). It is worth noting that the Cu $K_\alpha$ and $K_\beta$ X-rays are of greater intensities than the Ag $K_\alpha$ and $K_\beta$ X-rays, respectively, due to the bremsstrahlung produced by the Mini-X having a much greater intensity at energies capable of exciting Cu K-shell electrons than at energies capable of exciting Ag K-shell electrons. Furthermore, two peaks at approximately 9.4 and 11.0 keV were observed. These peaks are at energies consistent with Au $L_\alpha$ and $L_\beta$ X-rays that have been Compton scattered through an angle of approximately 160°. The Au $L_\alpha$ and $L_\beta$ X-rays (with initial energies of 9.7 and 11.4 keV, respectively) were emitted by the Au Mini-X anode as the result of electron impact before being Compton scattered by the poltinnik. Compton-scattered bremsstrahlung from the Mini-X can also be observed in the spectrum; however, the continuous nature of the energy-distribution of the bremsstrahlung makes it difficult to discern it from background radiation.

The XRF spectrum produced by the poltinnik minted in 1921 is shown in Figure 3. The intensities of the Ag $K_\alpha$ and $K_\beta$ X-rays in the spectra are comparable to those produced by the 1913 poltinnik, suggesting that the Ag-contents of the coins are approximately the same. The intensities of the Cu K-shell X-rays in the spectrum suggest that the poltinnik struck in 1921 contains slightly less Cu than the 1913 poltinnik. Other than the aforementioned Compton-scattered Au L-shell X-rays, no other prominent peaks can be seen in the spectrum. It should be noted, however, that X-rays with energies less than 5 keV were not likely to be detected (due to poor detector efficiency) and the bremsstrahlung incident on the poltinniks could not result in the emission of characteristic X-rays with energies greater than 40 keV. Thus, if any of the poltinniks contain elements which emit characteristic radiation with energies below approximately 5 keV or above 40 keV, the presence of these elements may not have been detected in our experiments.

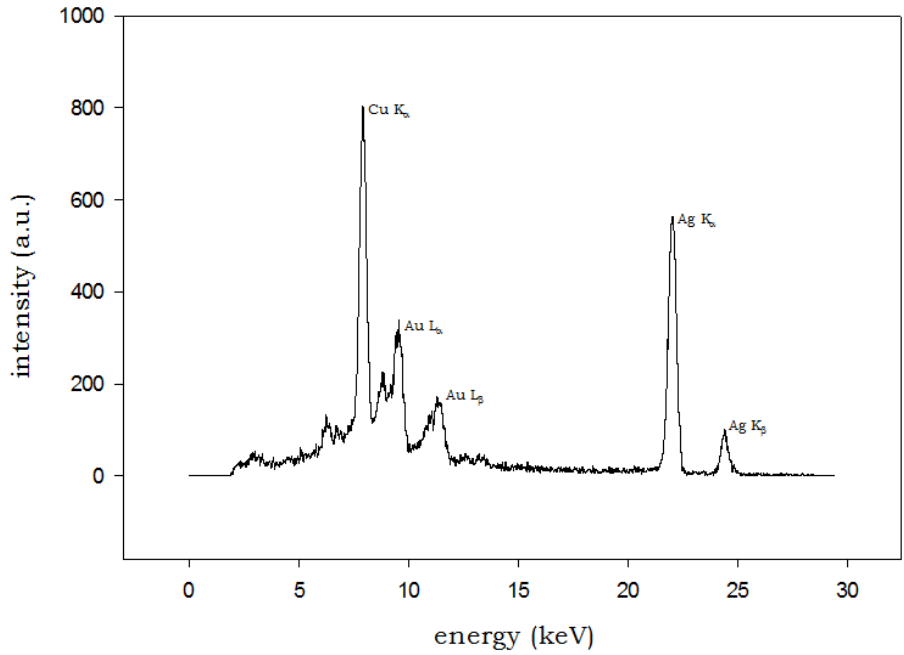

**Figure 2.** XRF spectrum produced by the 1913 poltinnik.

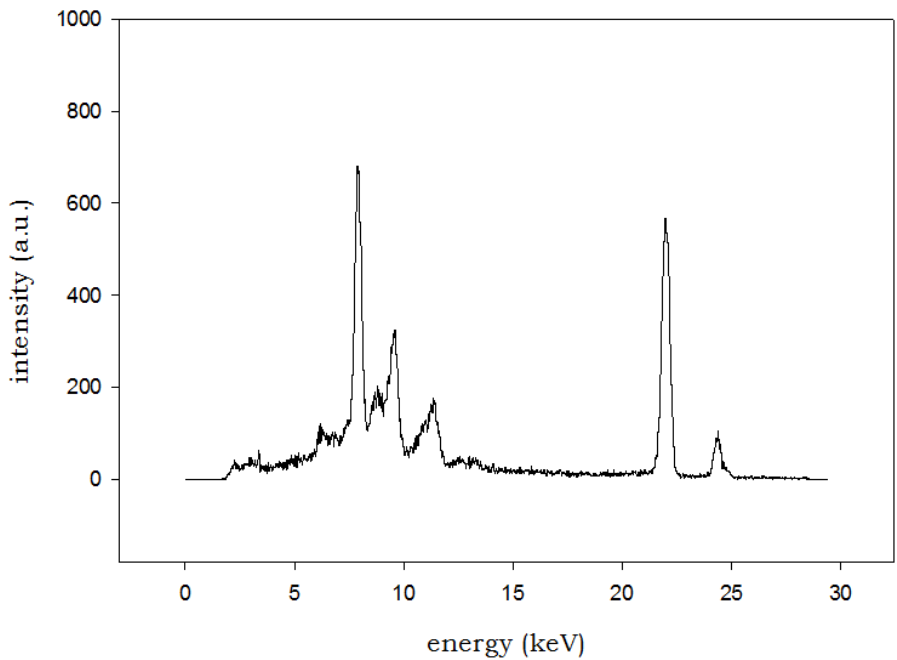

**Figure 3.** XRF spectrum produced by the 1921 poltinnik.

Figures 4 and 5 are the XRF spectra produced by the 1924 poltinniks minted in Leningrad and London, respectively. The spectra appear essentially identical to one another, suggesting that there is no appreciable difference in the Ag or Cu content of the coins. Furthermore, a comparison of Figures 4 and 5 to the XRF spectra associated with the 1913 and 1921 poltinniks suggests that the amounts of Ag and Cu used in the minting of all of the coins were approximately equal and that the Decree of February 22, 1924 was carried out.

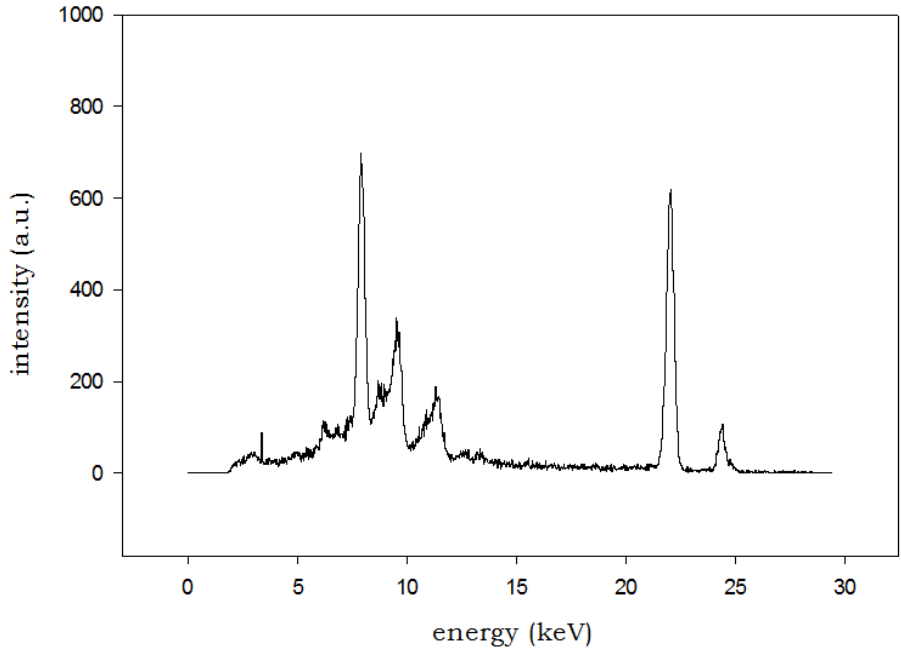

**Figure 4.** XRF spectrum produced by the 1924 poltinnik minted in Leningrad.

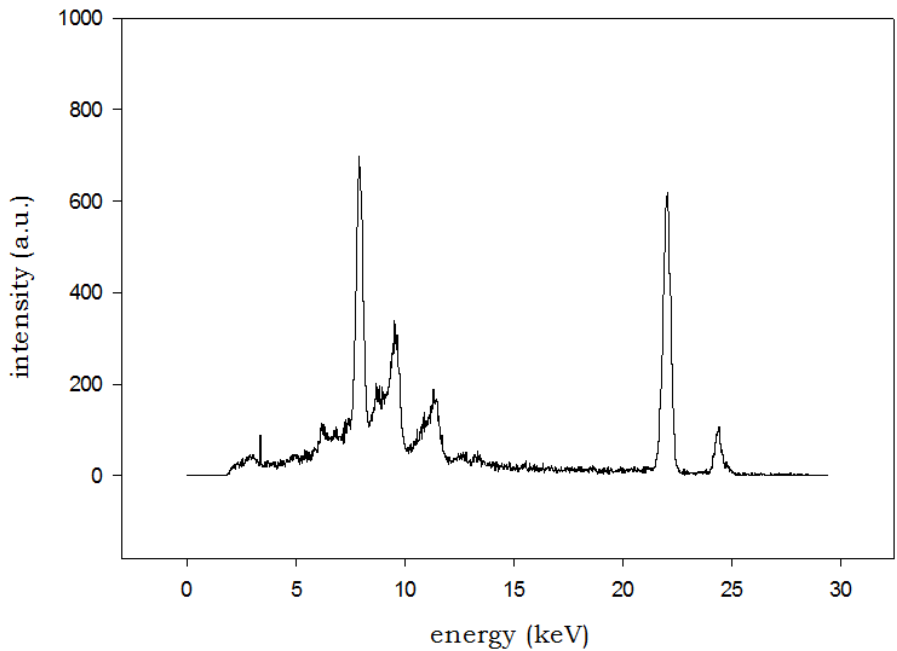

**Figure 5.** XRF spectrum produced by the 1924 poltinnik minted in London.

Table 1 compares the number of net counts in the region of the Ag $K_\alpha$ peaks in each of the spectra as well as the masses of each of the poltinniks. Background radiation was subtracted by averaging the number of counts in the ten channels on either side of the Ag $K_\alpha$ X-ray peak region, determining the average number of background counts per channel, multiplying this number by the number of channels across which the $K_\alpha$ peak is spread, and subtracting this value from the total number of counts in the region of the peak. The uncertainties given in the table are based solely on statistical error. Despite the fact that a cursory examination of Figures 2-5 suggests that the intensities of the Ag $K_\alpha$ X-rays were essentially equal, the data in Table 1 indicates that there were slight differences. A comparison of the net counts in the region of the Ag $K_\alpha$ peaks suggests that the 1924 poltinnik minted in Leningrad contains a slightly higher proportion of Ag than the others, and that the 1924 poltinnik struck in London contains a slightly lower proportion of Ag than the others. The percent difference between the net counts in the region of the Ag $K_\alpha$ peaks in the spectra produced by the 1924 poltinniks is slightly less than 5.5%. As previously mentioned, the surfaces of the coins were not treated prior to XRF analysis. When purchased, the 1924 poltinnik minted in Leningrad appeared to have been professionally cleaned (as opposed to the others, which did not appear to have had their surfaces treated). It may be that surface treatment is responsible for the high intensity of the Ag $K_\alpha$ X-rays emitted by the coin, as XRF is essentially a surface analysis technique [3]. The masses of all of the coins were $9.9 \pm 0.15$ g, suggesting that the portion of the Decree of February 22, 1924 pertaining to the weights of the poltinniks was carried out  Any differences in mass may be attributed to wear or irregularities in the minting processes.

| poltinnik | net counts | uncertainty | mass +/- 0.05 (grams) |
|---|---|---|---|
| 1913 | 8578 | 93 | 9.9 |
| 1921 | 8784 | 94 | 9.8 |
| 1924 (ПЛ) | 8842 | 94 | 10.0 |
| 1924 (TP) | 8372 | 91 | 9.9 |

**Table 1.** Comparison of the net counts (and uncertainties) in the region of the Ag $K_\alpha$ peaks in each of the spectra, as well as the masses of each of the poltinniks. "TP" and "ПЛ" are the mintmarks on the 1924 poltinniks minted in London and Leningrad, respectively.

### 4. Conclusions

We analyzed four poltinniks minted in 1913, 1921, and 1924 using XRF as part of an undergraduate research project. The fact that both the Ag-content of each of the poltinniks and the masses of the coins are essentially the same leads us to conclude that the Decree of February 22, 1924 was carried out. Simple XRF experiments, such as the ones described here and in previous reports [7, 8], can give undergraduate students both valuable "hands-on" experience working with equipment commonly found in many laboratories and a better understanding of several radiative processes including characteristic X-ray emission, Compton scattering, and bremsstrahlung emission.

**Acknowledgement**

The authors wish to thank N. McGara for his help setting up some of the equipment used in our experiments.